\documentclass[11pt,twoside]{article}
\usepackage[pdftex]{graphicx}
\usepackage{amsmath}
\usepackage{amssymb}
\usepackage{cite}
\begin{document}

\newcommand{\pst}{\hspace*{1.5em}}

\newcommand{\rigmark}{\em Journal of Russian Laser Research}
\newcommand{\lemark}{\em Volume 30, Number 5, 2009}

\newcommand{\be}{\begin{equation}}
\newcommand{\ee}{\end{equation}}
\newcommand{\bm}{\boldmath}
\newcommand{\ds}{\displaystyle}
\newcommand{\bea}{\begin{eqnarray}}
\newcommand{\eea}{\end{eqnarray}}
\newcommand{\ba}{\begin{array}}
\newcommand{\ea}{\end{array}}
\newcommand{\arcsinh}{\mathop{\rm arcsinh}\nolimits}
\newcommand{\arctanh}{\mathop{\rm arctanh}\nolimits}
\newcommand{\bc}{\begin{center}}
\newcommand{\ec}{\end{center}}

\thispagestyle{plain}

\label{sh}


\begin{center} {\Large \bf
\begin{tabular}{c}
 CENTER OF MASS TOMOGRAPHY AND WIGNER FUNCTION 
 \\[-1mm]
 FOR MULTIMODE PHOTON STATES.
\end{tabular}
 } \end{center}

\bigskip

\bigskip

\begin{center} {\bf
Ivan V. Dudinets$^{1*}$ and Vladimir I. Man'ko $^2$
}\end{center}

\begin{center}
{\it
$^1$Moscow Institute of Physics and Technology (State University)\\
Institutskii per. 9, Dolgoprudnii, Moscow Region 141700, Russia

\smallskip

$^2$P.N. Lebedev Physical Institute, Russian Academy of Sciences\\
Leninskii Prospect 53, Moscow 119991, Russia
}
\smallskip

$^*$Corresponding author e-mail: dudinets@phystech.edu \\
\end{center}

\begin{abstract}\noindent
Tomographic probability representation of multimode electromagnetic field states in the scheme of center-of-mass tomography is reviewed. Both connection of the field state Wigner function and observable Weyl symbols with the center-of-mass tomograms as well as connection of Gr\"{o}newold kernel with the center-of-mass tomographic kernel determining the noncommutative product of the tomograms are obtained. The dual center-of-mass tomogram of the photon states are constructed and the dual tomographic kernel is obtained. The models of other generalised center-of-mass tomographies are discussed. Example of two-mode Schr{\"o}dinger cat states is presented in details.  
\end{abstract}

\medskip


\section{Introduction}
\pst
There exists tomographic probability representations of quantum states~\cite{Ibort2009, ManciniTombesi96}. Among these representations the optical tomography scheme based on relations of the optical tomogram with the Wigner function~\cite{Wigner32} discussed in~\cite{BerBer,VogelRisken} as well as the symplectic tomography scheme introduced in~\cite{ManciniTombesi96,ManciniTombesi97}. The spin tomography was constructed in~\cite{DodonovMan97,MankoVIandOV97,D'arianoPaini}. The center-of-mass tomography was introduced in~\cite{Arkhipov_LasRes} and developed in~\cite{ArkhipovPRA}. The review of tomographic picture of quantum mechanics is presented in~\cite{Ibort2009,MarmoVentPhysScr2015}. All the tomographic schemes provide description of quantum states in term of fair probability distributions. The  tomograms of the states are connected by integral transforms with quasiprobability distributions like the Wigner function, Husimi Q-function~\cite{Husimi40} and Glauber-Sudarshan P-function~\cite{Sudar1963P-func,Glauber1963}. The aim of our work is to study in details the center-of-mass tomographic probability representations of multimode electromagnetic field states. The tomographic approach can be presented in terms of the quantizer-dequantizer formalism~\cite{Marmo_JPhysA2002}. In the work we use this formalism to find the relation of the Wigner function, Weyl symbols of observables and integral kernels determining the star-product of the observable symbols e.g. Gr\"{o}newold kernel~\cite{Gronewald46} with the corresponding center-of-mass tomograms in the case of multimode electromagnetic field states. 
	The paper is organised as follows.
	In Sec.~2 we review the quantizer-dequantizer formalism (star-product formalism). In Sec.~3 we present the contribution to the center-of-mass tomography. The dual center-of-mass tomography is considered and explicit form of the star-product integral kernel of two-mode center-of-mass tomographic symbols is obtained. Connection between the Weyl correspondence and the center-of-mass map is given in Sec.~4. Some other modifications of the center-of-mass tomography are developed in Sec.~5. An ambiguity in the center-of-mass tomographic description of quantum states is discussed in Sec.~6. Example of superpositions of two-mode coherent states is gived in Sec.~7. Prospectives and conclusions are formulated in Sec.~8.
	
\section{The star-product scheme}
\pst
In this section we review the star-product formalism following~\cite{Marmo_JPhysA2002}.
In quantum mechanics, physical observables are represented by operators acting in a given Hilbert space $\mathcal{H}$. According to the star-product formalism one can construct an invertible map of  operators onto  functions. Thus one can use functions instead of operators. The invertible map can be constructed with the aid of families of operators  quantizers $\hat{D}(\mathbf{x})$ and dequantizers $\hat{U}(\mathbf{x})$ labelled by a vector with $n$ components $\mathbf{x}=(x_1,x_2,\ldots,x_n)$.
Given an operator $\hat{A}$ acting in  $\mathcal{H}$ the corresponding function (called 'symbol' of the operator $\hat{A}$) is defined by the formula
\be\label{dequant}
w_A(\mathbf{x})=\mbox{Tr}[\hat{A}\,\hat{U}(\mathbf{x})].
\ee 
The formula 
\be\label{quant}
\hat{A}=\int w_A(\mathbf{x})\hat{D}(\mathbf{x})d\mathbf{x}
\ee
allows to reconstruct the operator from its simbol. In the latter formula an integration over  continuous and sum over discreate components of the vector $\mathbf{x}$ are assumed. Let us notice that formulae~(\ref{dequant}) and~(\ref{quant}) are compatible if for the symbol $w_A(\mathbf{x})$ of any operator $\hat{A}$ the following identity holds true 
\be\label{compcond0}
w_A(\mathbf{x})=\int w_A(\mathbf{x'})\mbox{Tr}[\hat{D}(\mathbf{x'})\hat{U}(\mathbf{x})]d\mathbf{x'}.
\ee
Deriving the above formula, we assumed that one can exchange the trace with the integral. Let $w_A(\mathbf{x})$ and $w_B(\mathbf{x})$ be symbols of operators $\hat{A}$ and $\hat{B}$, then 
for the operator $\hat{A}\hat{B}$ corresponding symbol is
\be
w_{AB}(\mathbf{x})=\mbox{Tr}[\hat{A}\,\hat{B}\,\hat{U}(\mathbf{x})]=\int w_A(\mathbf{x}_2)w_B(\mathbf{x}_1)\mbox{Tr}[\hat{D}(\mathbf{x}_2)\hat{D}(\mathbf{x}_1)\hat{U}(\mathbf{x})]d\mathbf{x}_1 d\mathbf{x}_2.
\ee
The symbol $w_{AB}(\mathbf{x})$ is called the star-product of symbols   $w_{A}(\mathbf{x})$ and $w_{B}(\mathbf{x})$ and denoted $(w_A\star w_B)(\mathbf{x})$ and the expression
\be\label{kernel}
K(\mathbf{x}_2,\mathbf{x}_1,\mathbf{x})=\mbox{Tr}[\hat{D}(\mathbf{x}_2)\hat{D}(\mathbf{x}_1)\hat{U}(\mathbf{x})]
\ee
called the kernel of star-product~\cite{Marmo_JPhysA2002}. Since the standard product of operators is associative, i.e. $\hat{A}(\hat{B}\hat{C})=(\hat{A}\hat{B})\hat{C}$, the star-product of symbols of operators must be associative too
\be\label{associativ.cond}
w_A(\mathbf{x})\star (w_B\, \star w_C)(\mathbf{x})=(w_A\star w_B)(\mathbf{x})\star w_C(\mathbf{x}).  
\ee
The associativity condition~(\ref{associativ.cond}) in terms of the kernel  of star-product of symbols of operators takes the form~\cite{MarmoVitalePhysLett2007}
\be\label{associativ.cond.kernel}
\int K(\mathbf{x}_1,\mathbf{x}_2,\mathbf{y})K(\mathbf{y},\mathbf{x}_3,\mathbf{x}_4)d\mathbf{y}=\int K(\mathbf{x}_1,\mathbf{y},\mathbf{x}_4)K(\mathbf{x}_2,\mathbf{x}_3,\mathbf{y})d\mathbf{y}.
\ee
Let us suppose that there exists another families  quantizers $\hat{D'}(\mathbf{y})$ and dequantizers $\hat{U'}(\mathbf{y})$ acting in $\mathcal{H}$ and labelled by a vector with $m$ components $\mathbf{y}=(y_1,y_2,\ldots,y_m)$. Using these operators, for an operator $\hat{A}$ one can associate another function different from~(\ref{dequant})  
\be\label{dequantphi}
w'_A(\mathbf{y})=\mbox{Tr}[\hat{A}\,\hat{U'}(\mathbf{y})],
\ee 
the inverse relation is
\be\label{quantphi}
\hat{A}=\int w'_A(\mathbf{y})\hat{D}'(\mathbf{y})d\mathbf{y}.
\ee
Since the functions $w_A(\mathbf{x})$ and $w'_A(\mathbf{y})$ are symbols of the same operator $\hat{A}$, one can obtain by inserting~(\ref{quant}) into~(\ref{dequantphi})~\cite{Marmo_JPhysA2002}
\be\label{phif}
w'_A(\mathbf{y})=\int w_A(\mathbf{x})\mbox{Tr}[\hat{D}(\mathbf{x})\hat{U}'(\mathbf{y})]d\mathbf{x}.
\ee
The last formula provides the relation between symbols corresponding to different maps.
Similarly, using formulae~(\ref{dequant}) and~(\ref{quantphi}), one gets the inverse relation 
\be\label{fphi}
w_A(\mathbf{x})=\int w'_A(\mathbf{y})\mbox{Tr}[\hat{D}'(\mathbf{y})\hat{U}(\mathbf{x})]d\mathbf{y}.
\ee

Using the definition of the star-product kernel~(\ref{kernel}), one gets that
the kernels corresponding to different maps are connected to each other by the following relation
\be\label{conn_kernels}
K'(\mathbf{y}_1,\mathbf{y}_2,\mathbf{y}_3)=\int K(\mathbf{x}_1,\mathbf{x}_2,\mathbf{x}_3)\mbox{Tr}[\hat{D}'(\mathbf{y}_1)\hat{U}(\mathbf{x}_1)]\mbox{Tr}[\hat{D}'(\mathbf{y}_2)\hat{U}(\mathbf{x}_2)]\mbox{Tr}[\hat{U}'(\mathbf{y}_3)\hat{D}(\mathbf{x}_3)]d\mathbf{x}_1d\mathbf{x}_2d\mathbf{x}_3,
\ee
where $K'(\mathbf{y}_2,\mathbf{y}_1,\mathbf{y})=\mbox{Tr}[\hat{D}'(\mathbf{y}_2)\hat{D}'(\mathbf{y}_1)\hat{U}'(\mathbf{y})]$.

In~\cite{MarmoVitalePhysLett2005} the special case of the map was considered
\be\label{dualexchang}
\hat{U}^d(\mathbf{x})=\hat{D}(\mathbf{x}),\quad \hat{D}^d(\mathbf{x})=\hat{U}(\mathbf{x}),
\ee
where authors have exchanged the dequantizer $\hat{U}(\mathbf{x})$ and the quantizer $\hat{D}(\mathbf{x})$. The new pair quantizer-dequantizer is called dual to the initial one. 
Let us define the dual symbol of an operator $\hat{A}$ 
\be\label{dual}
w^{d}_A(\mathbf{x})=\mbox{Tr}[\hat{A}\,\hat{U^d}(\mathbf{x})]=\mbox{Tr}[\hat{A}
\,\hat{D}(\mathbf{x})].
\ee 
The reconstruction formula for the operator $\hat{A}$ is given by 
\be\label{dualinverse}
\hat{A}=\int w^{d}_A(\mathbf{x})\hat{D^d}(\mathbf{x})d\mathbf{x}=\int w^{d}_A(\mathbf{x})\hat{U}(\mathbf{x})d\mathbf{x}.
\ee
The dual operators provide a new associated star-product with the kernel
\be\label{dualkernel}
K^d(\mathbf{x}_2,\mathbf{x}_1,\mathbf{x})=\mbox{Tr}[\hat{D^d}(\mathbf{x}_2)\hat{D^d}(\mathbf{x}_1)\hat{U^d}(\mathbf{x})]=\mbox{Tr}[\hat{U}(\mathbf{x}_2)\hat{U}(\mathbf{x}_1)\hat{D}(\mathbf{x})].
\ee
The important property of the dual map is that the mean value of an observable $\hat{A}$ is given by the product of the symbol of the density operator and the symbol of the observable in the dual representation 
\be\label{mean}
\langle \hat{A}\rangle = \mbox{Tr} \hat{\rho}\hat{A}=\mbox{Tr}\left(\hat{A}\int w_{\rho}(\mathbf{x})\hat{D}(\mathbf{x})\, d\mathbf{x}\right)=\int w_{\rho}(\mathbf{x})\mbox{Tr}\left(\hat{A}\hat{D}(\mathbf{x})\right)\, d\mathbf{x}=
\int w_{\rho}(\mathbf{x})w^d_{A}(\mathbf{x})\, d\mathbf{x},
\ee
where $w_{\rho}$ is the symbol of the density operator $\hat{\rho}$, namely $w_{\rho}(\mathbf{x})=\mbox{Tr}[\hat{\rho}\,\hat{U}(\mathbf{x})]$
(see Eq.~(\ref{dequant})) and $w^d_{A}$ is the symbol of an observable $\hat{A}$ in dual representation.

\section{The symplectic tomographic and the center-of-mass maps}
In this section we consider some special cases of tomographic maps, namely the symplectic, the center-of-mass and the dual center-of-mass maps. Here and throughout the paper we regard a quantum  system with $N$ degrees of freedom (for example, $N=md$ for $m$ particles in $d$ dimension). Each vector has $N$ components if otherwise stated. In the case of the symplectic map one choose $\mathbf{x}=(\vec{X},\vec{\mu},\vec{\nu})$. 
The quantizer and the dequantizer for the symplectic map are given by the formula

\be\label{Us}
\hat{U}_{s}(\vec{X},\vec{\mu},\vec{\nu})=\delta(\vec{X}-\vec{\mu}\hat{\vec{q}}-\vec{\nu}\hat{\vec{p}}),
\ee
\be\label{Ds}
\hat{D}_{s}(\vec{X},\vec{\mu},\vec{\nu})=(2\pi)^{-N}e^{i(\vec{e}\vec{X}-\vec{\mu}\hat{\vec{q}}-\vec{\nu}\hat{\vec{p}})},
\ee
where $\hat{\vec{q}}$ and $\hat{\vec{p}}$ are the vectors with components $\hat{q}_j$ and $\hat{p}_j$ being position and momentum operators for each degree of freedom, the vector $\vec{e}$ has all components equal to 1 and $\vec{a}\,\vec{b}$ denotes scalar product of two vectors $\vec{a}$ and $\vec{b}$. The vector $\vec{X}$ can be associated with the positions of the system in scaled and rotated reference frame in phase space, $\vec{\mu}$ and $\vec{\nu}$ being real parameters of scaling and rotation. For any operator one can associate the tomographic symbol according to~(\ref{dequant}) and~(\ref{Us}). By definition, the symplectic tomogram is the symbol of the density operator
\be\label{symplTom}
w_{s}(\vec{X},\vec{\mu},\vec{\nu})=\mbox{Tr}[\hat{\rho}\,\delta(\vec{X}-\vec{\mu}\hat{\vec{q}}-\vec{\nu}\hat{\vec{p}})].
\ee
According to formula~(\ref{quant}), the density operator can be reconstructed from the symplectic tomogram
\be\label{rhosimpl.tom}
\hat{\rho}=(2\pi)^{-N}\int w_{s}(\vec{X},\vec{\mu},\vec{\nu}) e^{i(\vec{e}\vec{X}-\vec{\mu}\hat{\vec{q}}-\vec{\nu}\hat{\vec{p}})} d\vec{X} d\vec{\mu}\, d\vec{\nu}.
\ee
The state of a system with $N$ degrees of freedom can be described by the density matrix of $2N$ variables. In view of~(\ref{symplTom}) and~(\ref{rhosimpl.tom}), the same state can be either described by the symplectic tomogram, which is the nonnegative function of $3N$ variables, which is less convenient for large $N$. This due to the fact that the symplectic tomogram has extra variables, which do not give additional information about the quantum system.  
However, in~\cite{ArkhipovPRA} authors constructed the map called the center-of-mass map which allows to circumvent this problem. In the case of the center-of-mass map $\mathbf{x}=(X,\vec{\mu},\vec{\nu})$, vectors $\vec{\mu}$ and $\vec{\nu}$ are with $N$ components each and $X$ is real. The quantizer and the dequantizer are of the form
\be\label{cmU}
\hat{U}_{cm}(X,\vec{\mu},\vec{\nu})=\delta(X-\vec{\mu}\hat{\vec{q}}-\vec{\nu}\hat{\vec{p}}),
\ee
\be\label{cmD}
\hat{D}_{cm}(X,\vec{\mu},\vec{\nu})=(2\pi)^{-N}e^{i(X-\vec{\mu}\hat{\vec{q}}-\vec{\nu}\hat{\vec{p}})}.
\ee
The dequantizer~(\ref{cmU}) and the quantizer~(\ref{cmD}) determine the center-of-mass kernel of star-product
\be\label{cm_kernel}
K_{cm}(X_1,\vec{\mu}_1,\vec{\nu}_1, X_2,\vec{\mu}_2,\vec{\nu}_2, X_3,\vec{\mu}_3,\vec{\nu}_3)=\frac{e^{i(X_1+X_2)+\frac{i}{2}(\vec{\nu}_1\vec{\mu}_2-\vec{\mu}_1\vec{\nu}_2)}}{(2\pi)^{N+1}}\int e^{ik X_3}\delta(\vec{\mu}_1+\vec{\mu}_2+k\vec{\mu}_3)\,\delta(\vec{\nu}_1+\vec{\nu}_2+k\vec{\nu}_3)\,dk.
\ee
In the latter formula the integration of $2N$ delta-functions is readily performed. For instance, in the case of two degrees of freedom $\vec{\mu}=(\mu^{(1)},\mu^{(2)})$, $\vec{\mu_i}=(\mu_i^{(1)},\mu_i^{(2)})$, $i=1,2$ and the similar formulae for $\vec{\nu}$, $\vec{\nu_i}$ the center-of-mass kernel has the form
\be\nonumber
K_{cm}(X_1,\vec{\mu}_1,\vec{\nu}_1, X_2,\vec{\mu}_2,\vec{\nu}_2, X_3,\vec{\mu}_3,\vec{\nu}_3)=\frac{e^{i (X_1+X_2)-iX_3\frac{\nu^{(2)}_1+\nu^{(2)}_2}{\nu^{(2)}_3}
+\frac{i}{2}(\nu^{(1)}_1 \mu^{(1)}_2-\mu^{(1)}_1\nu^{(1)}_2+\nu^{(2)}_1 \mu^{(2)}_2-\mu^{(2)}_1\nu^{(2)}_2)}}{(2\pi)^{3}|\nu^{(1)}_3\nu^{(2)}_3\mu^{(1)}_3\mu^{(2)}_3|}
\ee
\be\label{cm_two}
\times \delta \left ( \frac{\mu^{(1)}_1+\mu^{(1)}_2}{\mu^{(1)}_3}-\frac{\nu^{(2)}_1+\nu^{(2)}_2}{\nu^{(2)}_3} \right )\delta \left ( \frac{\mu^{(2)}_1+\mu^{(2)}_2}{\mu^{(2)}_3}-\frac{\nu^{(2)}_1+\nu^{(2)}_2}{\nu^{(2)}_3} \right )
\delta \left ( \frac{\nu^{(1)}_1+\nu^{(1)}_2}{\nu^{(1)}_3}-\frac{\nu^{(2)}_1+\nu^{(2)}_2}{\nu^{(2)}_3}  \right ).
\ee
The center-of-mass tomogram is defined as the symbol of the density operator 
\be\label{cmtomogram}
w_{cm}(X,\vec{\mu},\vec{\nu})=\mbox{Tr}[\hat{\rho}\,\delta(X-\vec{\mu}\hat{\vec{q}}-\vec{\nu}\hat{\vec{p}})].
\ee
According to formula~(\ref{quant}), the density operator can be reconstructed from the center-of-mass tomogram
\be\label{invertmap}
\hat{\rho}=(2\pi)^{-N}\int w_{cm}(X,\vec{\mu},\vec{\nu}) e^{i(X-\vec{\mu}\hat{\vec{q}}-\vec{\nu}\hat{\vec{p}})} dX d\vec{\mu}\, d\vec{\nu}.
\ee
Formulae~(\ref{cmtomogram}) and~(\ref{invertmap}) determine the invertible map between  the tomogram $w_{cm}(X,\vec{\mu},\vec{\nu})$ and the density operator of the system. Therefore, the quantum state of a system with $N$ degrees of freedom can be described by the nonnegative function with $2N+1$ variables. Furthermore, the center-of-mass tomogram is a homogeneous function, namely  $w_{cm}(\lambda X,\lambda \vec{\mu},\lambda \vec{\nu})=|\lambda|^{-1} w_{cm}(X,\vec{\mu},\vec{\nu})$ for any real $\lambda \neq 0$, which follows from~(\ref{cmtomogram}). Hence, the center-of-mass tomogram actually operates with $2N$ variables as the density matrix does. However, unlike the latter, the center-of-mass tomogram is nonegative function. Given the wave function of a pure system, the center-of-mass tomogram is determined by fractional Fourier transform of the wave function~\cite{ArkhipovPRA}.

According to general scheme~(\ref{phif}) and~(\ref{fphi}), the transition kernels between the
 symplectic and the center-of-mass maps read
\be
\mbox{Tr}[\hat{D}_{cm}(X_1,\vec{\mu}_1,\vec{\nu}_1)\hat{U}_{s}(\vec{X}_2,\vec{\mu}_2,\vec{\nu}_2)]=(2\pi)^{-N}e^{iX_1}\int e^{-i \vec{k}\vec{X}_2}\delta(\vec{\mu}_1-\vec{k}\circ\vec{\mu}_2)\,\delta(\vec{\nu}_1-\vec{k}\circ\vec{\nu}_2)\,d\vec{k},
\ee

\be
\mbox{Tr}[\hat{D}_{s}(\vec{X}_1,\vec{\mu}_1,\vec{\nu}_1)\hat{U}_{cm}(X_2,\vec{\mu}_2,\vec{\nu}_2)]=(2\pi)^{-1}e^{i\vec{e}\vec{X_1}}\int e^{-i k X_2}\delta(\vec{\mu}_1-k\vec{\mu}_2)\,\delta(\vec{\nu}_1-k \vec{\nu}_2)\,d k.
\ee
Here $\vec{a}\circ \vec{b}$ denotes the component-wise product of vectors $\vec{a}=(a_1,a_2,\ldots,a_N)$ and $\vec{b}=(b_1,b_2,\ldots,b_N)$, i.e. the vector $(a_1b_1,a_2b_2,\ldots,a_N b_N)$.
These kernels determine the relation between symbols in the symplectic and the center-of-mass representations
\be\label{sym_cm}
w_{s}(\vec{X},\vec{\mu},\vec{\nu})=(2\pi)^{-N}\int w_{cm}(Y,\vec{k}\circ\vec{\mu},\vec{k}\circ\vec{\nu})e^{i(Y-\vec{k}\vec{X})}d\vec{k}dY
\ee 
and
\be\label{cm_sym}
w_{cm}(X,\vec{\mu},\vec{\nu})=\int w_{s}(\vec{Y},\vec{\mu},\vec{\nu})\delta(X-\vec{e}\vec{Y})d\vec{Y}.
\ee 

In derivation of the last formula, we have used the homogeneity property of the symplectic tomogram $w_{s}(\lambda \vec{X},\lambda\vec{\mu},\lambda\vec{\nu})=|\lambda|^{-N}w_{s}(\vec{X},\vec{\mu},\vec{\nu})$, which is directly follows from~(\ref{symplTom}) and the homogeneity property of delta-function.

Exchanging the role of the quantizer~(\ref{cmU}) and the dequantizer~(\ref{cmD}), i.e.
\be
\hat{U}^d_{cm}(X,\vec{\mu},\vec{\nu})=(2\pi)^{-N}e^{i(X-\vec{\mu}\hat{\vec{q}}-\vec{\nu}\hat{\vec{p}})},
\ee
\be
\hat{D}^d_{cm}(X,\vec{\mu},\vec{\nu})=\delta(X-\vec{\mu}\hat{\vec{q}}-\vec{\nu}\hat{\vec{p}}),
\ee
one obtains the symbol of an operator $\hat{A}$ in the dual center-of-mass representation
\be
 w^d_{A}(X,\vec{\mu},\vec{\nu})=(2\pi)^{-N}e^{iX}\mbox{Tr}\left[\hat{A}\,e^{-i(\vec{\mu}\hat{\vec{q}}+\vec{\nu}\hat{\vec{p}})}\right].
\ee
The reconstruction formula provides an expression for the operator $\hat{A}$ in terms of its dual symbol
\be
 \hat{A}= \int w^d_{A}(X,\vec{\mu},\vec{\nu})\delta(X-\vec{\mu}\hat{\vec{q}}-\vec{\nu}\hat{\vec{p}})\,d\vec{\mu}\,d\vec{\nu}=
  (2\pi)^{-N}\int e^{i(\vec{\mu}\hat{\vec{q}}+\vec{\nu}\hat{\vec{p}})} \mbox{Tr}\left[\hat{A}\,e^{-i(\vec{\mu}\hat{\vec{q}}+\vec{\nu}\hat{\vec{p}})}\right]\,d\vec{\mu}\,d\vec{\nu}.
\ee
The dual kernel for the center-of-mass scheme determined by~(\ref{dualkernel}) is expressed as follows
\be
K^{d}_{cm}(X_2,\vec{\mu}_2,\vec{\nu}_2,X_1,\vec{\mu}_1,\vec{\nu}_1,X,\vec{\mu},\vec{\nu})=\int e^{iX-i\frac{k_1k_2}{2}(\vec{\mu}_2\vec{\nu}_1-\vec{\nu}_2\vec{\mu}_1)}
 e^{i(k_1 X_2+k_2 X_1)}\delta(k_1\vec{\mu}_2+k_2\vec{\mu}_1+\vec{\mu})\delta(k_1\vec{\nu}_2+k_2\vec{\nu}_1+\vec{\nu})\frac{dk_1 dk_2}{4\pi^2}.
\ee
According to the general rule~(\ref{mean}), the mean value of a quantum observable $\hat{A}$ is given by integration of the product of its dual symbol and the center-of-mass tomogram
\be
\langle \hat{A}\rangle =
\int w_{cm}(X,\vec{\mu},\vec{\nu})w^d_{A}(X,\vec{\mu},\vec{\nu})\, dX\,d\vec{\mu}\,d\vec{\nu}.
\ee
Since the center-of mass tomogram has the property of a fair probability distribution, the dual symbol $w^d_{A}(X,\vec{\mu},\vec{\nu})$ of an observable $\hat{A}$ plays the role of the function identified with the observable in
the center-of-mass scheme.

\section{Weyl correspondence}
In this section we present the general scheme that relates the center-of-mass tomographic map with the Weyl correspondence 
providing an invertible map of operators onto functions on phase space (Weyl symbols). 
The Weyl correspondence is the particular case of the star-product scheme and can be described by using the following pair of the dequantizer and the quantizer
\be\label{DU_phase}
\hat{U}(\vec{q},\vec{p})=\int e^{-i\vec{p}\vec{u}}\left|\vec{q}-\frac{\vec{u}}{2}\right \rangle \left \langle \vec{q}+\frac{\vec{u}}{2}\right|d\vec{u},\quad \hat{D}(\vec{q},\vec{p})=(2\pi)^{-N} \hat{U}(\vec{q},\vec{p}).
\ee
The Weyl symbol of an operator $\hat{A}$ is defined as follows

\be
w_{A}(\vec{q},\vec{p})=\mbox{Tr}\left[ \hat{A}\,\hat{U}(\vec{q},\vec{p})\right]=
\int e^{-i\vec{p}\vec{u}}\left \langle \vec{q}+\frac{\vec{u}}{2}\right| 
\hat{A} \left|\vec{q}-\frac{\vec{u}}{2}\right \rangle d\vec{u}.
\ee
The operator $\hat{A}$ can be reconstructed from its Weyl symbol
\be
\hat{A}=\int w_{A}(\vec{q},\vec{p}) \hat{D}(\vec{q},\vec{p}) d\vec{q} d\vec{p}.
\ee
The Wigner function is defined as the Weyl symbol of the density operator, i.e. 
\be
W(\vec{q},\vec{p})=\int e^{-i\vec{p}\vec{u}}\left \langle \vec{q}+\frac{\vec{u}}{2}\right| 
\hat{\rho} \left|\vec{q}-\frac{\vec{u}}{2}\right \rangle d\vec{u}.
\ee
The relation between symbols of the center-of-mass map and the Weyl correspondence is given by Eqs.~(\ref{phif})-(\ref{fphi})
\be
w_{A}(X,\vec{\mu},\vec{\nu})=\int w_A(\vec{q},\vec{p}) \delta(X-\vec{\mu}\vec{q}-\vec{\nu}\vec{p})d\vec{q}d\vec{p}.
\ee
and
\be
w_A(\vec{q},\vec{p})=\int w_{A}(X,\vec{\mu},\vec{\nu}) e^{i(X-\vec{\mu}\vec{q}-\vec{\nu}\vec{p})}dX d\vec{\mu}d\vec{\nu}.
\ee
For the case of the density operator, i.e. $\hat{A}=\hat{\rho}$, one obtains the relation between the center-of-mass tomogram and the Wigner function
\be\label{Wig_cm}
w_{cm}(X,\vec{\mu},\vec{\nu})=\int W(\vec{q},\vec{p}) \delta(X-\vec{\mu}\vec{q}-\vec{\nu}\vec{p})d\vec{q}d\vec{p}.
\ee
The star-product of Weyl symbols 
\be\label{star_Weyl}
(w_A\star w_B)(\vec{q}_3,\vec{p}_3)=\int G(\vec{q}_1,\vec{p}_1,\vec{q}_2,\vec{p}_2,\vec{q}_3,\vec{p}_3) 
w_A(\vec{q}_1,\vec{p}_1)w_B(\vec{q}_2,\vec{p}_2)d\vec{q}_1\,d\vec{q}_2\,d\vec{p}_1\,d\vec{p}_2.
\ee
is determined by the Gr{\"o}newold kernel 
\be\label{Gron}
G(\vec{q}_1,\vec{p}_1,\vec{q}_2,\vec{p}_2,\vec{q}_3,\vec{p}_3)=\pi^{-2N}\exp 2i\left(
\vec{q}_1\vec{p}_2-\vec{q}_2\vec{p}_1+\vec{q}_2\vec{p}_3-\vec{q}_3\vec{p}_2+\vec{q}_3\vec{p}_1-\vec{q}_1\vec{p}_3
\right).
\ee
Using Eq.~(\ref{conn_kernels}) one obtains the relation between the Gr{\"o}newold and the center-of-mass kernels
\be\nonumber
K_{cm}(X_1,\vec{\mu}_1,\vec{\nu}_1, X_2,\vec{\mu}_2,\vec{\nu}_2, X_3,\vec{\mu}_3,\vec{\nu}_3)=(2\pi)^{-3N}e^{i(X_1+X_2)}
\int G(\vec{q}_1,\vec{p}_1,\vec{q}_2,\vec{p}_2,\vec{q}_3,\vec{p}_3)e^{-i(\vec{\mu}_1\vec{q}_1+\mu_2\vec{q}_2+\vec{\nu}_1\vec{p}_1+\vec{\nu}_2\vec{p}_2)}\ee
\be\label{cm_green}
\times\delta (X_3-\vec{\mu}_3\vec{q}_3-\vec{\nu}_3\vec{p}_3)d\vec{q}_1d\vec{p}_1d\vec{q}_2d\vec{p}_2d\vec{q}_3d\vec{p}_3.
\ee
In the standard classical mechanics formalism multiplication of functions on phase-space is given by the pointwise commutative and associative product
\be
w_A(\vec{q},\vec{p})\cdot w_B(\vec{q},\vec{p})=\int w_A(\vec{q}_1,\vec{p}_1) w_B(\vec{q}_2,\vec{p}_2)
\delta (\vec{q}-\vec{q}_1)\delta (\vec{q}-\vec{q}_2)\delta (\vec{p}-\vec{p}_1)\delta (\vec{p}-\vec{p}_2)
d\vec{q}_1\,d\vec{q}_2\,d\vec{p}_1\,d\vec{p}_2,
\ee
where the kernel reads 
\be\label{classic}
K_{cl}(\vec{q}_1,\vec{p}_1,\vec{q}_2,\vec{p}_2,\vec{q},\vec{p})=\delta (\vec{q}-\vec{q}_1)\delta (\vec{q}-\vec{q}_2)\delta (\vec{p}-\vec{p}_1)\delta (\vec{p}-\vec{p}_2).
\ee
It was shown in~\cite{ibort2016quantum} that the kernel of  the pointwise product is the limit $\hbar\rightarrow 0$ of the Gr{\"o}newold kernel with the Planck constant reinserted. Thus, in quantum mechanics the star-product of functions on phase-space  is determined by the Gr{\"o}newold kernel, whereas in classical mechanics functions on phase-space are multiplied according to the pointwise product. It worth noting that for two particles the kernel~(\ref{cm_two}), where the term in the exponent $\mu^{(1)}_1\nu^{(1)}_2-\nu^{(1)}_1\mu^{(1)}_2+\mu^{(2)}_1\nu^{(2)}_2-\nu^{(2)}_1\mu^{(2)}_2$ is removed corresponds to the point-wise product of functions on  phase-space. This statement can be proven by inserting~(\ref{classic}) into~(\ref{cm_green}) and taking the integrals.

\section{Cluster tomogram}

One can generalize the scheme of the center-of-mass map. To do that let us consider a quantum system with $N$ degrees of freedom composed of $r$ subsystems with $k$th subsystem having $N_k$ degrees of freedom (of course, the following equality holds $N=N_1+N_2+\ldots +N_r$). For each subsystem we construct the dequantizer and the quantizer of center-of-mass map~(\ref{cmU}) and~(\ref{cmD}), namely
$\hat{U}_k=\delta\left(X_k- \vec{\mu}_k\,\hat{\vec{q}}_k+\vec{\nu}_k\,\hat{\vec{p}}_k\right )$ and   $\hat{D}_{k}=(2\pi)^{-N_k}\exp{\left(X_k- \vec{\mu}_k\,\hat{\vec{q}}_k+\vec{\nu}_k\,\hat{\vec{p}}_k\right )},$ where $\vec{\mu}_k$ and  $\vec{\nu}_k$ are $N_k$-components vectors with entries $\mu_k^{(i)}$ and $\nu_k^{(i)}$, $i=1,\ldots,N_k$, $\hat{\vec{q}}_k$ and $\hat{\vec{p}}_k$ are $N_k$-components vectors with entries $\hat{q}_k^{(i)}$  and $\hat{p}_k^{(i)}$ being position and momentum operators for $k$th subsystem. Here $X_k$ is the sum of the positions of $k$th subsystem measured in rotated and scaled reference frame in phase space, $\mu_k^{(i)}$ and $\nu_k^{(i)}$ being the real parameters of scaling and rotation. Let us introduce $r$-components vector $\vec{X}=(X_1,X_2,\ldots,X_r)$  and $N$-components vectors $\vec{\mu}=(\vec{\mu}_1,\vec{\mu}_2,\ldots, \vec{\mu}_r) $, $\vec{\nu}=(\vec{\nu}_1,\vec{\nu}_2, \ldots, \vec{\nu}_r)$. With the composed system (cluster) we associate the dequantizer and the quantizer as product of the dequantizer and the quantizer of each subsystem, respectively 
\be\label{Ucl}
\hat{U}_{cl}(\vec{X},\vec{\mu},\vec{\nu})=\prod_{k=1}^{r}\delta\left(X_k- \vec{\mu}_k\,\hat{\vec{q}}_k+\vec{\nu}_k\,\hat{\vec{p}}_k\right ),
\ee
\be\label{Dcl}
\hat{D}_{cl}(\vec{X},\vec{\mu},\vec{\nu})=(2\pi)^{-N}\prod_{k=1}^{r}\exp{\left(X_k- \vec{\mu}_k\,\hat{\vec{q}}_k+\vec{\nu}_k\,\hat{\vec{p}}_k\right )}. 
\ee
It is worth noting that the cases $r=1$ and $r=N$ correspond to the center-of-mass and the symplectic maps, respectively. The Kernel corresponding to the quantizer and the  dequantizer~(\ref{Dcl}),~(\ref{Ucl}) reads
\be\nonumber
K_{cl}(\vec{X}'',\vec{\mu}'',\vec{\nu}'',\vec{X}',\vec{\mu}',\vec{\nu}',\vec{X},\vec{\mu},\vec{\nu})=
(2\pi)^{-N-r}e^{i \vec{e}\,\left(\vec{X}''+\vec{X}'\right) +i ( \vec{\mu}^{'}\vec{\nu}^{''}-\vec{\nu}^{'}\vec{\mu}^{''})/2} 
\ee
\be
\times\int d\vec{k}\,e^{-i\vec{k} \vec{X}  }
\delta\left ( \vec{\mu}^{''}+\vec{\mu}^{'}-\vec{k}\circ\vec{\mu} \right )\delta\left ( \vec{\nu}^{''}+\vec{\nu}^{'}-\vec{k}\circ\vec{\nu} \right ),
\ee
where $\vec{k}$ and $\vec{e}$ are $r$-components vectors.
The cluster tomogram of the composed system with the density operator $\hat{\rho}$ is defined by 
\be
w_{cl}(\vec{X},\vec{\mu},\vec{\nu})=
\mbox{Tr}\left[ \hat{\rho}\, \hat{U}_{cl}(\vec{X},\vec{\mu},\vec{\nu})\right]=
\mbox{Tr}\left[ \hat{\rho}\, \prod_{k=1}^{r}\delta\left(X_k- \vec{\mu}_k\,\hat{\vec{q}}_k+\vec{\nu}_k\,\hat{\vec{p}}_k\right )\right].
\ee
The tomogram $w_{cl}(\vec{X},\vec{\mu},\vec{\nu})$ is a nonnegative normalized function 
\be
\int w_{cl}(\vec{X},\vec{\mu},\vec{\nu}) d\vec{X}=1.
\ee
The connection between the center-of-mass and the cluster tomogram reads
\be\label{cl_cm}
w_{cl}(\vec{X},\vec{\mu},\vec{\nu})=(2\pi)^{-r}\int w_{cm}(Y,\vec{k}\circ\vec{\mu},\vec{k}\circ\vec{\nu})e^{i(Y-\vec{k}\vec{X})}d\vec{k}dY
\ee 
and
\be\label{cm_cl}
w_{cm}(X,\vec{\mu},\vec{\nu})=\int w_{cl}(\vec{Y},\vec{\mu},\vec{\nu})\delta(X-\vec{e}\vec{Y})d\vec{Y}.
\ee 
The state of $m$th subsystem is associated with the center-of-mass tomogram
\be\label{sub_cl}
w_{cm}^{(m)}(X_m,\vec{\mu}_m,\vec{\nu}_m )=\int w_{cl}(\vec{X},\vec{\mu},\vec{\nu}) dX_1 \ldots dX_{m-1}dX_{m+1}\ldots dX_{r}.
\ee
Using~(\ref{cl_cm}), one can obtain the expression for the center-of-mass tomogram of $m$th subsystem in terms of the center-of-mass tomogram of the composed system
\be\label{sub_cm}
w_{cm}^{(m)}(X_m,\vec{\mu}_m,\vec{\nu}_m )=(2\pi)^{-1}\int w_{cm}\left( Y,k \vec{a}_m,k\vec{b}_m \right) e^{i(Y-kX_m)}dk\,dY,
\ee
where $\vec{a}_m=(0,\ldots,0,\vec{\mu}_m,0,\ldots,0)$ and $\vec{b}_m=(0,\ldots,0,\vec{\nu}_m,0,\ldots,0)$ are $N$-components vectors with the first $N_1+N_2+\ldots+N_{m-1}$ components being zero. 

As an example, let us consider a system with $N$ degrees of freedom composed of two subsystems having $N_1$ and $N_2$ degrees of freedom ($N=N_1+N_2$). The state of the composed system can be described both by the cluster tomogram $w_{cl}(\vec{X},\vec{\mu},\vec{\nu})$ and by the center-of-mass tomogram $w_{cm}(X,\vec{\mu},\vec{\nu})$, where 
$\vec{\mu}=(\vec{\mu}_1,\vec{\mu}_2)$, $\vec{\nu}=(\vec{\nu}_1,\vec{\nu}_2)$ are $N$-components vectors, $\vec{X}=(X_1,X_2)$. The state of the first subsystem is associated with the center-of-mass tomogram~(\ref{sub_cl}) 
\be
w_{cm}^{(1)}(X_1,\vec{\mu}_1,\vec{\nu}_1 )=\int w_{cl}(\vec{X},\vec{\mu},\vec{\nu}) dX_2. 
\ee
The expression for $w_{cm}^{(1)}$ in terms of the center-of-mass tomogram of the composed system reads
\be\label{cmonemode}
w_{cm}^{(1)}(X_1,\vec{\mu}_1,\vec{\nu}_1 )=(2\pi)^{-1}\int w_{cm}\left(Y,k \vec{a}_1, k\vec{b}_1 \right) e^{i(Y-kX_1)}dk\,dY, 
\ee
where $\vec{a}_1=(\vec{\mu}_1,0,\ldots,0)$ and  $\vec{b}_1=(\vec{\nu}_1,0,\ldots,0)$.

The simplest case of the cluster tomogram corresponds to the  factorized density operator, i.e
$\hat{\rho}=\hat{\rho}_1\otimes \hat{\rho}_2\otimes\ldots \otimes\hat{\rho}_r$
\be
 w_{cl}(\vec{X},\vec{\mu},\vec{\nu})= \prod_{m=1}^{r}w_{cl}^{(m)}(X_m,\vec{\mu}^{(m)},\vec{\nu}^{(m)} ).
\ee
Thus, the cluster tomogram for the systems without correlations reduced to the products of the center-of-mass tomogram of each subsystems.

\section{Joint probability distribution for the center-of-mass tomogram}

It was pointed out in the previous section that the states of quantum systems can be identified with the center-of-mass tomograms being the probability distributions of a random continuous variable $X$ and extra parameters $\mu_j$ and $\nu_j$. It was indicated in~\cite{man2012tomographic} that the center-of-mass tomogram can be treated as conditional probability distribution, and corresponding notation is 
\be w_{cm}(X|\vec{\mu},\vec{\nu})\equiv w_{cm}(X,\vec{\mu},\vec{\nu}).\ee

This interpretation follows from the fact that the center-of-mass tomogram satisfies so-called 'no-signalling' property (see, e.g.,~\cite{man2012tomographic}).
\be\label{normaliz}
\int w_{cm}(X|\vec{\mu},\vec{\nu})\,dX =1, 
\ee
which holds true for any parameters $\vec{\mu}$ and $\vec{\nu}$. For the center-of-mass tomogram one can construct a joint probability of random variables $X$, $\vec{\mu}$ and $\vec{\nu}$ in view of the Bayes formula (see, e.g.,~\cite{ holevo2011probabilistic})
\be
\mathcal{W}(X,\vec{\mu},\vec{\nu})= w_{cm}(X|\vec{\mu},\vec{\nu})P(\vec{\mu},\vec{\nu}),
\ee
where $P(\vec{\mu},\vec{\nu})$ is an arbitrary nonnegative normalized function  $\int P(\vec{\mu},\vec{\nu})d\vec{\mu} d\vec{\nu}=1$. For example, one can take  the Gaussian distribution function
 $P(\vec{\mu},\vec{\nu})= \pi^{-N}\exp{\left(-\vec{\mu}^2-\vec{\nu}^2\right)}$, where $\vec{\mu}^2$ denotes the the usual square of a vector $\vec{\mu}$, i.e. $\vec{\mu}^2=\sum \mu_i^2$. It is obviously that 
\be
\int \mathcal{W}(X,\vec{\mu},\vec{\nu})dX = P(\vec{\mu},\vec{\nu}).
\ee
The nonnegative function $\mathcal{W}(X,\vec{\mu},\vec{\nu})$ is normalized  with respect
to all the variables, i.e. 
\be
\int \mathcal{W}(X,\vec{\mu},\vec{\nu})dXd\vec{\mu} d\vec{\nu}=1.
\ee
Inversely, given a joint probability distribution $\mathcal{W}(X,\vec{\mu},\vec{\nu})$ one can introduce the condition probability function 
\be
w_{cm}(X|\vec{\mu},\vec{\nu})=\mathcal{W}(X,\vec{\mu},\vec{\nu})\left( \int \mathcal{W}(X,\vec{\mu},\vec{\nu})dX\right)^{-1},
\ee
which satisfies the no-signalling property.

According to~(\ref{invertmap}), the density matrix can be expressed in terms of the joint probability distribution
\be
\hat{\rho}=\int \mathcal{W}(X,\vec{\mu},\vec{\nu})\left( \int \mathcal{W}(X,\vec{\mu},\vec{\nu})dX\right)^{-1}\exp{[i(X-\vec{\mu}\hat{\vec{q}}-\vec{\nu}\hat{\vec{p}})]} \frac{dX d\vec{\mu}\, d\vec{\nu}}{(2\pi)^N}.
\ee
It follows from the latter formula that states of quantum systems can be associated with the joint probability distributions. Thus, there exist an ambiguity in constructing such probability distributions, which is related to the choice of the distribution function of random parameters.

\section{Center-of-mass tomogram of the two-mode Schr{\"o}dinger cat states}

Let us consider a system with two one-dimensional subsystems. The state of the composed system $\hat{\rho}_{12}$ can be described both by the symplectic tomogram $w_s(X_1,X_2,\mu_1,\mu_2,\nu_1,\nu_2)$ and by the center-of-mass tomogram $w_{cm}(X,\mu_1,\mu_2,\nu_1,\nu_2)$. The expression for the density operator $\hat{\rho}_{12}$ in terms of the tomogram $w_{cm}$ is given by~(\ref{invertmap}) 
\be
\hat{\rho}_{12}=(2\pi)^{-2}\int w_{cm}(X,\mu_1,\mu_2,\nu_1,\nu_2) e^{i(X-\mu_1\hat{q}_1-\mu_2\hat{q}_2-\nu_1\hat{p}_1-\nu_2\hat{p}_2)} dX d\mu_1\, d\mu_2\, d\nu_1 \,d\nu_2,
\ee
where $\hat{q}_i$, $\hat{p}_i$ are position and momentum operators for $i$th subsystem, $i=1,2$.
The density operator of the first subsystem can be obtained by performing a trace over the second subsystem
\be\label{1subsrho}
\hat{\rho}_{1}=\mbox{Tr}_2\,\hat{\rho}_{12}=(2\pi)^{-1}\int w_{cm}(X,\mu_1,0,\nu_1,0) e^{i(X-\mu_1\hat{q}_1-\nu_1\hat{p}_1)} dX d\mu_1\, d\nu_1.
\ee
The symplectic tomogram of the first subsystem reads 
\be\label{1subscm}
w_1(X,\mu_1,\nu_1)=\mbox{Tr}\hat{\rho}_1\delta(X-\mu_1\hat{q}_1-\nu_1\hat{p}_1)=
(2\pi)^{-1}\int w_{cm}(X,k\mu_1,0,k\nu_1,0) e^{i(X-k X_1)} dk\,dX.
\ee
The latter formula corresponds to $N_1=N_2=1$ in~(\ref{cmonemode}).

\begin{figure}[ht]
\begin{center}
\begin{minipage}[h]{0.49\linewidth}
\includegraphics[width=1\linewidth]{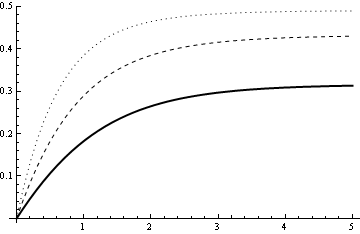}
\caption{The linear entropy $S_{+}(\alpha_1,\alpha_2)$~(\ref{cat_entrlin}) for $|\alpha_2|^2=0.5$ (solid line), $|\alpha_2|^2=1$ (dashed line), $|\alpha_2|^2=2$ (dotted line) and various values of $|\alpha_1|^2$.}
\label{fig:linpl}
\end{minipage}
\hfill 
\begin{minipage}[h]{0.49\linewidth}
\includegraphics[width=1\linewidth]{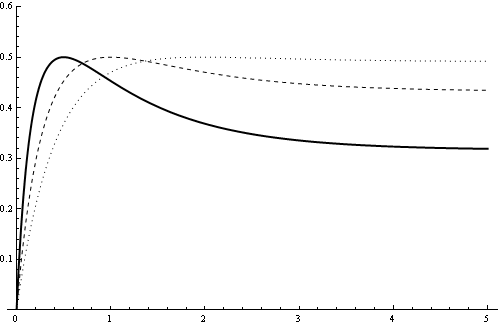}
\caption{The linear entropy $S_{-}(\alpha_1,\alpha_2)$~(\ref{cat_entrlin}) for $|\alpha_2|^2=0.5$ (solid line), $|\alpha_2|^2=1$ (dashed line), $|\alpha_2|^2=2$ (dotted line) and various values of $|\alpha_1|^2$.}
\label{fig:linmin}
\end{minipage}
\end{center}
\end{figure}

Let us suppose that the composed system is entangled. As a measure of entanglement, we use the linear entropy defined as 
\be\label{linearentr}
S_1=1-\mbox{Tr}\hat{\rho}^2_1.
\ee 
The linear entropy ranges from $1$, corresponding to a separable state, to $0$ for a maximally entangled state.
Inserting~(\ref{1subsrho}) into~(\ref{linearentr}), one gets the expression for the linear entropy  in terms of the center-of-mass tomogram 
\be\label{entr_cm}
S_1=1-(2\pi)^{-1}\int w_{cm}(X,\mu,0,\nu,0)w_{cm}(Y,-\mu,0,-\nu,0) e^{i(X+Y)} d\mu\,d\nu\,dXdY. 
\ee
As an example, let us consider the two-mode Schr{\"o}dinger cat states, which are even and odd superpositions of coherent states with opposite phases~\cite{DodonovMalkinManKo1974}
\be\label{schr_cat}
\psi_{\pm}(x_1,x_2)=N_{\pm}(\alpha)(\psi_{\alpha}(x_1,x_2)\pm\psi_{-\alpha}(x_1,x_2)).
\ee
The normalization constant is given by $N_{\pm}^{-2}(\alpha)=(2\pm 2 e^{-2|\alpha_1|^2-2|\alpha_2|^2}) $.
Here $\psi_{\alpha}(x_1,x_2)$ is the wave function of two-mode coherent state $|\alpha_1\rangle |\alpha_2\rangle$
labeled by complex vector $\vec{\alpha}=(\alpha_1,\alpha_2)$ 
\be
\psi_{\alpha}(x_1,x_2)=\pi^{-N/4}\exp \left(-\frac{x^2_1+x^2_2}{2}+\sqrt{2}(\alpha_1 x_1+\alpha_2 x_2) -\frac{|\alpha_1|^2+|\alpha_2|^2}{2}-\frac{\alpha^2_1+\alpha^2_2}{2}\right).
\ee
Note that the states $\psi_{\pm}(x_1,x_2)$ are entangled.
The center-of-mass tomogram for the Schr{\"o}dinger cat states can be calculated by means of the Wigner function (see Eq.~(\ref{Wig_cm})). Omitting the straightforward calculations, we obtain    
\be\nonumber
w_{cm,\alpha}(X,\mu_1,\mu_2,\nu_1,\nu_2)=\pi^{-1/2}\sigma^{-1/2}N_{\pm}^{2}(\alpha)(
\exp{\left((X-\sqrt{2}\,\mathcal{R}\alpha_1\,\mu_1-\sqrt{2}\,\mathcal{R}\alpha_2\,\mu_2-\sqrt{2}\,\mathcal{I}\alpha_1\,\nu_1-\sqrt{2}\,\mathcal{I}\alpha_2\,\nu_2)^2/ \sigma\right)}\ee
\be\nonumber 
\pm\exp{\left((X-i\sqrt{2}\,\mathcal{I}\alpha_1\,\mu_1-i\sqrt{2}\,\mathcal{I}\alpha_2\,\mu_2+i\sqrt{2}\,\mathcal{R}\alpha_1\,\nu_1+i\sqrt{2}\,\mathcal{R}\alpha_2\,\nu_2)^2/ \sigma\right)}\ee
\be\nonumber 
\pm\exp{\left((X+i\sqrt{2}\,\mathcal{I}\alpha_1\,\mu_1+i\sqrt{2}\,\mathcal{I}\alpha_2\,\mu_2-i\sqrt{2}\,\mathcal{R}\alpha_1\,\nu_1-i\sqrt{2}\,\mathcal{R}\alpha_2\,\nu_2)^2/ \sigma\right)}
\ee
\be\label{cm_cat}
\exp{\left((X+\sqrt{2}\,\mathcal{R}\alpha_1\,\mu_1+\sqrt{2}\,\mathcal{R}\alpha_2\,\mu_2+\sqrt{2}\,\mathcal{I}\alpha_1\,\nu_1+\sqrt{2}\,\mathcal{I}\alpha_2\,\nu_2)^2/ \sigma\right)}
),
\ee
where $\mathcal{R}\alpha$ and $\mathcal{I}\alpha$  denotes the real and the imaginary parts of complex variable $\alpha$ and $\sigma=\mu^2_1+\mu^2_2+\nu^2_1+\nu^2_2$. 
Inserting~(\ref{cm_cat}) into~(\ref{entr_cm}), we obtain the explicit expression for the linear entropy of the two-mode Schr{\"o}dinger cat states~(\ref{schr_cat})
\be\label{cat_entrlin}
S_{\pm}(\alpha_1,\alpha_2)=0.5-0.5\left( 1\pm e^{-2|\alpha_1|^2-2|\alpha_2|^2}\right)^{-2} \left(e^{-2|\alpha_1|^2}\pm e^{-2|\alpha_2|^2} \right)^2.
\ee 
The linear entropy $S_{\pm}(\alpha_1,\alpha_2)$ versus $|\alpha_1|^2$ and various values of  $|\alpha_2|^2$ is depicted in Figures~\ref{fig:linpl} and~\ref{fig:linmin}. The entropy $S_{+}(\alpha_1,\alpha_2)$ veries from $0$ for $|\alpha_1|=0$ to $0.5-0.5 e^{-4|\alpha_2|^2}$ for large values of $|\alpha_1|$. The entropy $S_{-}(\alpha_1,\alpha_2)$ increases from $0$ to 0.5 corresponding to $|\alpha_1|=|\alpha_2|$ and decreases to $0.5-0.5 e^{-4|\alpha_2|^2}$ for large values of $|\alpha_1|$.

\section{Conclusions}
To conclude, we point out the main results of our work.

In this article we have considered the center-of-mass map of operators onto functions (tomographic symbols) in the context of the star-product formalism given by a pair of quantizer - dequantizer operators. These functions depend on one random variable $X$ interpreted as "the center of mass" coordinate of the quantum system under consideration in rotated and scaled reference frame in phase space and extra real parameters $\vec{\mu}$ and $\vec{\nu}$.  The functions are multiplied according to a non-local and non-commutative product determined by the center-of-mass kernel (see~(\ref{cm_kernel})). We have obtained the connection between kernels of star-product corresponding to different maps, in particular we have given the relation between the center-of-mass and the Gr\"{o}newold kernels. We have studied the dual center-of-mass map and derived the kernel corresponding to the star-product of symbols of this map. 

The center-of-mass tomogram is defined as the function corresponding to the density operator of a quantum system. The center-of-mass tomogram being the probability distribution of $X$  determines the quantum state completely in the sense that given the center-of-mass tomogram one can obtain the density operator. We have obtained the connection between the center-of-mass tomogram and the Wigner function. We have discussed the generalization of the center-of-mass tomogram (the cluster tomogram). In view of the fact that the center-of-mass tomogram satisfies 'no-signalling' property, the tomogram can be considered as the conditional probability distribution of random variable $X$. The latter allows to construct a joint probability distribution of variables $X$, $\vec{\mu}$ and $\vec{\nu}$ with the help of an arbitrary nonnegative normalized function $P(\vec{\mu}, \vec{\nu})$, which gives rise to an ambiguity in the center-of-mass description of quantum states.

We have considered an example of two mode two-mode Schr{\"o}dinger cat states in details.

\bibliography{liter}

\end{document}